# OVERCOMING OBSTACLES TO INTEGRABILITY

## I: PERTURBED DIFFUSIVE SYSTEMS

Alex Veksler[1] and Yair Zarmi[1,2]
Ben-Gurion University of the Negev, Israel
[1]Department of Physics, Beer-Sheva, 84105
[2]Department of Energy & Environmental Physics
Jacob Blaustein Institute for Desert Research, Sede-Boqer Campus, 84990

ABSTRACT

Obstacles to integrability sometimes hamper the standard Normal Form (NF) analysis of perturbed integrable evolution equations. One is then forced to account for them by the Normal Form, which is the dynamical equation obeyed by the zero-order term. This spoils the integrability of the NF, and the simple wave-nature of the zero-order approximation. To avoid both undesired results, one must require that the obstacles be accounted for by the higher-order terms in the expansion of the solution (the Near-Identity Transformation – NIT). We show that this goal can be achieved if the higher-order terms are allowed to depend explicitly on the independent variables, $t$ and $x$ (an option that is not considered in the standard NF analysis). In addition, a particular algorithm for the construction of the NIT leads to a "canonical" form for the obstacles; they are expressed in terms of the symmetries of the unperturbed equation. The canonical form ensures the explicit vanishing of the obstacles in the case of a single-wave zero-order solution. In the case of a multiple-wave solution, the effect of the "canonical" obstacles on the higher-order terms in the NIT is confined to the region of interaction among the waves. This often facilitates the derivation of closed-form expressions for the asymptotic effect of the obstacles. These ideas are demonstrated for the cases of the perturbed Burgers and heat diffusion equations.

## I. Introduction

There is great interest in solutions of integrable nonlinear evolution equations [1], e.g., fronts (the Burgers equation [2, 7]), and solitons (the KdV [3-12], NLS [12-16] and other equations [17-22]). When a small perturbation is added, the resulting equations are often analyzed by the method of Normal Forms (NF) [23-27]. The motivation is the expectation that the NF, which is the dynamical equation for the zero-order approximation to the solution of the perturbed equation, will be also integrable and preserve the wave nature of the solution of the unperturbed equation. However, the analysis often leads to the emergence of *obstacles to integrability* [28-35]. These are terms that the formalism generates, which cannot be accounted for by the perturbative expansion of the solution (the Near Identity Transformation – NIT). Obstacles do not appear when the zero-order solution is a single wave, e.g. a front or a soliton; the NF then merely updates the dispersion relation obeyed by the wave velocity [23-35]. In the general case (e.g., multiple-wave solutions), except for specific forms of the perturbation, obstacles do emerge [31-35]. The only way to account for them in the standard NF analysis is to include them in the NF, rendering the latter non-integrable (hence the name "obstacles to integrability"), and, possibly, spoiling physical properties of the zero-order term (e.g., its wave structure). Obstacles to integrability are also encountered in the Lie-group analysis of perturbed evolution equations [33, 36].

The cause of these difficulties is the assumption, usually made in the NF analysis, that all the terms in the NIT are differential polynomials in $u$, the zero-order approximation, and do not depend explicitly on the independent variables ($t$ and $x$). We show that by giving up this assumption, the NIT can account for the obstacles. As a result, they need not be included in the NF, and, hence, cease to be obstacles to the integrability of the latter. The solution of the NF (the zero-order approximation) then retains the character of the solution of the unperturbed equation. We propose a specific algorithm for the construction of the higher-order terms in the NIT, which yields obstacles in a "canonical" form, expressible in terms of symmetries of the unperturbed equation. This form manifestly ensures that the obstacles vanish in the case of a single-wave solution. (Obstacles do not emerge in this case also in the standard NF analysis, however, there, this is not an obvious result.) For multi-wave solutions, the obstacles have a simple structure, which often leads to closed-form expressions for the asymptotic behavior of the higher-order terms in the NIT.

In [37], the results of these ideas for single- and two-wave solutions of the perturbed Burgers, heat-diffusion and KdV equations are summarized. In the present paper, we present the detailed analysis of the first two systems. Their main advantage is that obstacles emerge already in the first-order analysis [31,34], making the computation much simpler than in other equations of physical interest, e.g., the perturbed KdV or NLS equations, where obstacles to integrability emerge only in second-order [30,35]. Also, subsets of the obstacles vanish for the anti-symmetric configuration of equal and opposite wave numbers ($k_2 = -k_1$) of the two-wave solutions of the NF for both equations. Therefore, to analyze the general case ($k_2 \neq -k_1$) one can perform a Galilean transformation to the anti-symmetric configuration, where some or all of the obstacles vanish, carry out the NF analysis there, and then go back to the original problem by the inverse Galilean transformation. Although the perturbed heat diffusion equation is not studied often in the context of the subject matter of this paper, we discuss it for its pedagogical merit. In the standard NF analysis of this equation, obstacles emerge, which spoil the integrability of the NF and the wave nature of the zero-order solution. Allowing the correction terms in the NIT to depend explicitly on $t$ and $x$, the obstacles can be overcome, and their effect on the solution can be found in closed form, when the zero-order solution is a superposition of a finite number of waves. Finally, obstacles to integrability have been encountered in the Lie group analysis of the perturbed Burgers equation [31]. The NF analysis may help reveal the structure of the generators of a Lie transformation [38,39] that resolves the obstacle problem.

## 2. NF analysis of the perturbed Burgers equation

### 2.1 The unperturbed Burgers equation

The unperturbed equation

$$u_t = S_2[u] = 2\,u\,u_x + u_{xx}$$
(2.1)

is integrable. The solution that represents $n$ wave fronts is written in the form

$$u(t,x) = \tilde{u}(\xi_1, \xi_2, \ldots, \xi_n) = \frac{\sum_{i=1}^{n} k_i A_i \exp(k_i \xi_i)}{1 + \sum_{i=1}^{n} A_i \exp(k_i \xi_i)},$$
(2.2)

where

$$\xi_i = x - v_{i0}\,t, \qquad v_{i0} = -k_i, \qquad A_i > 0.$$
(2.3)

$S_n$, the *symmetries* of the unperturbed equation are defined for any solution by the recursion relation [40, 41, 43]

$$S_{n+1} = \partial_x\{S_n + u\,G_n\},$$
(2.4)

with

$$S_n[u] = \partial_x G_n[u], \qquad G_0 = 1.$$
(2.5)

Eq. (2.5) yields a relation that will turn out to be useful in the following:

$$S_n = G_{n+1} - G_1 G_n.$$
(2.6)

For the present analysis, we shall need

$$G_1 = u, \qquad S_1 = u_x$$
$$G_2 = u^2 + u_x, \qquad S_2 = 2uu_x + u_{xx}$$
$$G_3 = u^3 + 3uu_x + u_{xx}, \qquad S_3 = 3u^2 u_x + 3uu_{xx} + 3u_x^2 + u_{xxx} \quad . \tag{2.7}$$

The Lie brackets of any two symmetries vanish [8,11,38,39]:

$$\left[ S_m[u], S_n[u] \right] \equiv \sum_i \left\{ \frac{\partial S_m}{\partial u_i} \partial_x^i S_n - \frac{\partial S_n}{\partial u_i} \partial_x^i S_m \right\} = 0 \qquad \left( u_i \equiv \partial_x^i u \right) \quad . \tag{2.8}$$

*"Obstacles to integrability"* emerge in the analysis of the perturbed Burgers equation.  We will propose for them a specific "canonical" form, given by

$$R_{nm} = S_n[u] G_m[u] - S_m[u] G_n[u] = G_{n+1} G_m - G_{m+1} G_n \quad . \tag{2.9}$$

While playing an important role in the perturbative analysis to be discussed in the following sections, some $R_{nm}$ emerge already when one considers the time dependence of $G_n$.  in the study of the unperturbed equation.  Using Eqs. (2.4)-(2.6) one proves by induction that

$$\partial_t G_n = S_{n+1} + R_{n1} \quad . \tag{2.10}$$

## 2.2 Usual formal analysis leads to obstacles [31-35]

We first review of the manner in which obstacles emerge in the perturbed Burgers equation

$$w_t = 2ww_x + w_{xx} + \varepsilon \left( 3\alpha_1 w^2 w_x + 3\alpha_2 ww_{xx} + 3\alpha_3 w_x^2 + \alpha_4 w_{xxx} \right) +$$
$$\varepsilon^2 \left( 4\beta_1 w^3 w_x + 6\beta_2 w^2 w_{xx} + 12\beta_3 ww_x^2 + 4\beta_4 ww_{xxx} + 10\beta_5 w_x w_{xx} + \beta_6 w_{xxxx} \right) + O\left( \varepsilon^3 \right)$$
$$\left( \varepsilon \ll 1 \right) \quad . \tag{2.11}$$

We assume a power-series expansion (*Near Identity Transformation*, NIT) for $w$:

$$w = u + \varepsilon u^{(1)} + \varepsilon^2 u^{(2)} + \dots \quad . \tag{2.12}$$

The evolution of the zero-order term, $u(x, t)$, is governed by the *Normal Form* (NF), which is constructed from symmetries [39, 43]:

$$u_t = S_2[u] + \varepsilon\, \alpha_4\, S_3[u] + \varepsilon^2\, \beta_6\, S_4[u] + \dots \quad . \tag{2.13}$$

(To be consistent with the literature, we adhere to the usual expansion in terms of differential polynomials in $u$, i.e., functions of $u$, $u_x$, $u_{xx}$,..., although the same results are obtained at a lesser computational cost if the expansion is performed in terms of $G_n$, as independent entities.) Using Eqs. (2.12) and (2.13) in Eq. (2.11), one obtains the first-order *homological equation*, which determines $u^{(1)}$, the correction term that yields an $O(\varepsilon)$ approximate solution:

$$u_t^{(1)}[u;t,x] + \alpha_4\, S_3[u] = \left[S_2[u], u^{(1)}[u;t,x]\right] + 3\alpha_1\, u^2\, u_x + 3\alpha_2\, u\, u_{xx} + 3\alpha_3\, u_x^2 + \alpha_4\, u_{xxx} \quad . \tag{2.14}$$

In Eq. (2.14), the fact that, in general, $u^{(1)}$ may depend on $t$ and $x$ explicitly is exposed.

Assuming that $u^{(1)}$ is a differential polynomial in $u$, with no explicit dependence on $t$ and $x$ an obstacle to integrability emerges already in the first-order analysis. Under that assumption, the allowed structure of $u^{(1)}$ can be deduced in a variety of ways, the simplest one is by assigning *weights* [6, 8, 9] of 1 to $u$ and $\partial_x$ and of $2 -$ to $\partial_t$.. The result is [31, 34]

$$u^{(1)} = a\, u^2 + b\, q\, u_x + c\, u_x \qquad \left( q \equiv \int_{-\infty}^{x} u(t,x)dx \right) \quad . \tag{2.15}$$

The form of $u^{(1)}$ given by Eq. (2.15) cannot account for all the four independent differential monomials that appear on the r.h.s. of Eq. (2.14). $u_{xxx}$, is eliminated by $S_3[u]$ (see Eq. (2.7)), so that three monomials, $u^2u_x$, $uu_{xx}$ and $u_x^2$, remain. The Lie brackets of the last term in Eq. (2.15) with $S_2$ vanish. Consequently, two free parameters, $a$ and $b$, remain at our disposal. Substitution of Eq. (2.15) in Eq. (2.14) generates in Eq. (2.14) the following differential polynomial:

$$R^{(1)} = \left\{2\,a + b + \alpha_1 + \alpha_2 + 2\,\alpha_3 - 4\,\alpha_4\right\}u^2\,u_x + 2\left\{b - \alpha_1 + 2\,\alpha_2 + \alpha_3 - 2\,\alpha_4\right\}u\,u_{xx}$$
$$+ 2\left\{a + \alpha_1 - \tfrac{1}{2}\alpha_2 + \tfrac{1}{2}\alpha_3 - \alpha_4\right\}u_x^{\ 2} + \gamma_{21}\left(u^2\,u_x + u\,u_{xx} - u_x^{\ 2}\right) \qquad , \quad (2.16)$$

where

$$\gamma_{21} \equiv \left(2\,\alpha_1 - \alpha_2 - 2\,\alpha_3 + \alpha_4\right) \, . \tag{2.17}$$

Thus, with the ansatz that $u^{(1)}$ is a differential polynomial in $u$, given by Eq. (2.15), it is impossible to make $R^{(1)}$ vanish, unless $\gamma_{21} = 0$ [31,34]. Otherwise, any choice of $a$ and $b$ can only eliminate some of the monomials but not all. Hence, the NIT cannot account for all the monomials, and one is forced to include the obstacle, which is the remainder of $R^{(1)}$, in the NF, Eq. (2.13). As $R^{(1)}$ is not a symmetry (in particular, it does not contain the linear term $u_{xxx}$, which is part of $S_3$), it spoils the integrability of the NF, which may now have to be solved numerically, and its solution may lose the simple wave nature of the solution of the unperturbed equation.

As Eq. (2.16) indicates, the structure of the unaccountable part is not unique. This non-uniqueness will play a central role in the analysis in the following Sections. It depends on the algorithm adopted for canceling terms in Eq. (2.14). For instance, if one insists on eliminating as many monomials as the formalism allows, then one possible choice is

$$a = -\tfrac{3}{2}\left(\alpha_3 - \alpha_4\right) \qquad\qquad b = -\tfrac{3}{2}\left(\alpha_2 - \alpha_4\right) \quad , \tag{2.18}$$

$$R^{(1)} = \tfrac{3}{2}\,\gamma_{21}\,u^2\,u_x \, . \tag{2.19}$$

In contradiction with the statement made in Section 1, that no obstacles emerge in the case of a single-wave zero-order solution, the obstacle of Eq. (2.15) does not vanish in that case. The reason is that some of the steps leading to Eq. (2.18) are not possible in the single-wave case. In the following, we shall exploit the non-uniqueness in the structure of the unaccountable terms to propose an algorithm for their elimination. This algorithm will only generate obstacles that do vanish explicitly in the case of a single-wave zero-order solution.

## 2.3 General characteristics of single-wave solution of NF

A single-wave solution of the NF is of the form $u(t,x) = u(\xi = x - vt)$. Consider first the unperturbed equation, Eq. (2.1). If $u(\xi)$ is a solution of that equation, then the latter becomes

$$S_2[u] = 2u u_\xi + u_{\xi\xi} = -v_0 u_\xi \quad , \qquad (2.20)$$

where $v$ assumes the value $v_0$ of the unperturbed velocity. With the recursion relation, Eq. (2.4), Eq. (2.20) implies the following sequence of relations for the higher symmetries in the hierarchy:

$$\begin{aligned} S_n\big[u[\xi]\big] &= c_n S_1\big[u[\xi]\big] \quad \Rightarrow \quad G_n[u] = c_n u \\ c_{n+1} &= -\big(v_0 + u(\xi = -\infty)\big)c_n + u(\xi = -\infty)^n \end{aligned} \quad . \qquad (2.21)$$

For general boundary values at $\pm\infty$, the recursion relation for $c_n$ is derived by induction. In the case of a zero boundary value at $-\infty$, to which we adhere from now on, one obtains the well known value of $c_n = (-v_0)^n$ [26].

Eq. (2.21) holds also for the same $u(\xi)$, computed at a shifted point $\xi = x - vt$, where $v_0$ has been replaced by an updated velocity, $v$, because the definition of the symmetries $S_n$ involves spatial derivatives only. This observation has two important consequences. First, Eq. (2.21) leads to relations among spatial derivatives of different orders of $u$, reducing the number of independent monomials in the homological equation, so that no obstacles emerge in any order. Second, $u(\xi)$, becomes a solution of the NF, Eq. (2.13), with the following updated velocity:

$$v = v_0 - \varepsilon \alpha_4 v_0^2 + \varepsilon^2 \beta_6 v_0^3 + O(\varepsilon^3) \quad . \qquad (2.22)$$

## 2.4 Perturbed Burgers equation: Single-front solution of NF

The single-wave solution of the NF, Eq. (2.13) is a front, the $n = 1$ case of Eq. (2.2), given by:

$$u(t,x) = \frac{kA\exp(k(x-vt))}{1 + A\exp(k(x-vt))}, \qquad (A > 0) \tag{2.23}$$

with the unperturbed velocity, $v_0$, replaced by an updated value, $v$, given by Eq. (2.22).

No obstacles emerge in the perturbative analysis of this case. We denote first-order correction in the NIT, Eq. (2.12) by $u_s^{(1)}[u]$ (the subscript $s$ indicating the fact that the correction is computed with $u$ — a single-wave solution). It is found by solving Eq. (2.14) either by direct substitution of Eq. (2.23), or by exploiting the relations induced among the symmetries $S_n[u]$ (equivalently, relations among spatial derivatives of $u$) due to the fact that $u$ is a single-wave solution. With one free parameter, $c$, the form of $u_s^{(1)}[u]$ is:

$$u_s^{(1)}[u] = -\tfrac{1}{2}(2\alpha_1 - \alpha_2 + \alpha_3 - 2\alpha_4)u^2 + (\alpha_1 - 2\alpha_2 - \alpha_3 + 2\alpha_4)qu_x + c\,u_x \tag{2.24}$$

## 2.5 Overcoming the first-order obstacle: General case and two-front example

We have seen that obstacles to integrability are expected in the case of the general zero-order solution. As indicated in the introduction, this problem can be resolved by allowing the correction terms in the NIT to depend explicitly on $t$ and $x$. This approach will be adopted from here on.

Our main interest is in multi-wave zero-order solutions, which in the case of the Burgers equation are multiple fronts, an example of which, the two-front solution, will be studied later on. These solutions become well-separated fronts that tend asymptotically each to a single-front solution in a large portion of the $t$-$x$ plane. There are regions in the $t$-$x$ plane where the waves collide and lose their identity.  Motivated by the observation that obstacles to integrability vanish identically in the case of the single-front solution, our goal is to construct the NIT in such a manner that, in the case of multi-front solutions, the obstacles vanish at least asymptotically in that part of the *t-x* plane where the fronts are well separated.  This will be achieved by taking advantage of the freedom in the construction of the first-order correction, $u^{(1)}$, pointed out to in Section 2.2.  We write $u^{(1)}$ as

$$u^{(1)} = u_s^{(1)} + u_r^{(1)} \quad . \tag{2.25}$$

In Eq. (2.25), $u_s^{(1)}[u]$ is given by Eq. (2.24), which has been obtained in the analysis of the single-front case.  However, the structure of $u$ itself is *not* that of a single-front solution (because the latter does not lead to the emergence of obstacles).  Instead, it should be taken to be some general solution of the NF, Eq. (2.13), e.g., a multi-front solution..  To account for the obstacle by the NIT rather than by the NF, the correction term, $u_r^{(1)}$, is allowed to depend explicitly on the independent variables, *t* and *x*.  Eq. (2.14) is reduced to an equation for $u_r^{(1)}$:

$$\partial_t u_r^{(1)} = 2 u_x u_r^{(1)} + 2 u \partial_x u_r^{(1)} + \partial_x^2 u_r^{(1)} + R^{(1)} \quad . \tag{2.26}$$

The obstacle has the following form

$$R^{(1)} = \gamma_{21} R_{21} \quad , \tag{2.27}$$

where $R_{21}$, (see Eq. (2.9)), can be also written as

$$R_{21} = S_2[u]G_1[u] - S_1[u]G_2[u] = u^2 u_x + u u_{xx} - u_x^2 \quad . \tag{2.28}$$

Unless $\gamma_{21}$ (given by Eq. (2.17)) vanishes, $R^{(1)}$ seemingly constitutes an "obstacle".    In fact, it ceases to be an obstacle to integrability of the NF due to the presence of $u_r^{(1)}$.  Solving Eq. (2.26) for $u_r^{(1)}$, one accounts for the "unaccountable" terms, and the burden is alleviated from the NF.  The latter remains a sum of symmetries; hence, remains integrable, and *u*, the zero-order approximation of the perturbed equation retains the multiple-front nature of the solution of the unperturbed equation.

Expressing the obstacle in terms of symmetries and their integrals is an important consequence of the algorithm defined by Eq. (2.25).   In the case of a single-front solution of the NF, all symmetries are proportional to $S_1$ and their integrals – to $G_1 = u$ (see Section 2.3).   Hence, the obstacle vanishes explicitly in that case.   In the standard NF analysis, vanishing of obstacles in the single-wave case is not an explicit result, as exemplified by the discussion that leads to Eq. (2.19).

Moreover, the algorithm of Eq. (2.25) ensures that the obstacle also vanishes asymptotically in the case of a multi-front solution in that region of the $t$-$x$ plane where the fronts are well separated. This can be seen in two ways.  First, $u_s^{(1)}[u]$ cancels all the terms in the homological equation that persist in the single-front case.   Therefore, in a region in the plane where the solution tends asymptotically at an exponential rate to a single-front one, the homological equation tends asymptotically to its form in the single-front case, and, hence, the obstacle vanishes at an exponential rate.   Second, as the solution tends to  well-separated single fronts, the symmetries become proportional to $S_1$ except for exponentially small deviations.   Hence, again, the obstacle vanishes exponentially.

For a specific example, we turn to the two-front zero-order solution of the NF, Eq. (2.13), given by the $n = 2$ case of Eq. (2.2):

$$u(t,x) = \frac{k_1 A_1 \exp\big(k_1\big(x - v_1\, t\big)\big) + k_2 A_2 \exp\big(k_2\big(x - v_2\, t\big)\big)}{1 + A_1 \exp\big(k_1\big(x - v_1\, t\big)\big) + A_2 \exp\big(k_2\big(x - v_2\, t\big)\big)} \tag{2.29}$$

The NF updates the two velocities independently, according to Eq. (2.22).

One can check by substitution that $R_{21}$ of Eq. (2.28) does not vanish in the case of a two-front solution of the NF (Eq. (2.29)) but does vanish in the case of a single-front solution (Eq. (2.23)).

The price to be paid is that Eq. (2.26) can only be solved numerically. Still, the asymptotic effect of the obstacle on $u^{(1)}$ can be found in closed form. We note that, for $k_2 \cdot k_2 < 0$, the two fronts in the zero-order solution of Eq. (2.29) are distinct for $t << 0$. Near the origin in time they coalesce and form a single front which persists for all $t > 0$, centered around the line $x = -(k_1+k_2) \cdot t$ (see Fig. 1a). Thus, we interpret $t \geq 0$ as the interaction region.

Using Eq. (2.29) for $u$, the asymptotic behavior of the canonical obstacle $R_{21}$ is found to be (see Figs. 1b and 1c):

$$R_{21} \to \begin{cases} 0 & t \to -\infty \\ k_1 k_2 u_x & t \to +\infty \end{cases} .$$

(2.30)

Substituting Eq. (2.30) in Eq. (2.26), the asymptotic form of $u_r^{(1)}$ is found to be

$$u_r^{(1)} = \begin{cases} 0 & t << 0 \\ -\frac{1}{2}\gamma_{21} k_1 k_2, & t >> 0 \end{cases} .$$

(2.31)

The limits for $t > 0$ and $t < 0$ are interchanged for $k_2 \cdot k_2 > 0$. These results may be obtained in a number of ways (see the Appendix for one demonstration).

## 2.6 Overcoming the second-order obstacle: General case and two-front example

There is little interest in going to higher orders in the case of the perturbed Burgers equation, because the term proportional to $q(x,t)$, generated by the formalism already in the first-order term (see Eq. (2.24)), is unbounded (asymptotically it is linear in $x$). Therefore, it limits the validity of the approximation to $|t|$ and $|x|$ of $O(1)$. However, to demonstrate how our approach systematically generates obstacles of a similar structure in higher orders, we present in the following the results of the second-order analysis, which goes along the same steps as the first-order case. The $O(\varepsilon^2)$ homological equation has the form

$$
\begin{aligned}
u_t^{(2)}[u;t,x] + \beta_6\, S_4\,[u] = \\
\left[S_2[u]\, u^{(2)}[u;t,x]\right] + Z_2\left[u,u^{(1)}[u;t,x]\right] \\
+\, 4\beta_1\, u^3\, u_x + 6\beta_2\, u^2\, u_{xx} + 12\beta_3\, u\, u_x^2 + 4\beta_4\, u\, u_{xxx} + 10\beta_5\, u_x\, u_{xx} + \beta_6\, u_{xxxx} \quad .
\end{aligned}
\tag{2.32}
$$

As in Eq. (2.14), the fact that $u^{(1)}$ and $u^{(2)}$ may depend on $t$ and $x$ explicitly is exposed. $Z_2[u,u^{(1)}]$ is a term that the perturbative procedure generates from the known quantities, $u$ and $u^{(1)}$, which have been computed in the previous orders.

In analogy to Eq. (2.25), we write $u^{(2)}$ in the form

$$
u^{(2)} = u_s^{(2)}[u] + u_r^{(2)}(t,x) \quad .
\tag{2.33}
$$

The structure of the differential polynomial, $u_s^{(2)}[u]$, is found by solving Eq. (2.32) for $u$ that is a single-front solution (Eq. (2.23)). After $u_s^{(2)}[u]$ has been found, the single-wave $u$ is replaced by a solution that is not a single-wave one, e.g., the two-front solution of Eq. (2.29). $u_r^{(2)}$ is found by solving an equation analogous to Eq. (2.26):

$$
\partial_t u_r^{(2)} = 2\, u_x\, u_r^{(2)} + 2u\, \partial_x u_r^{(2)} + \partial_x{}^2 u_r^{(2)} + R^{(2)} \quad .
\tag{2.34}
$$

The full solution of Eq. (2.34) can only be found numerically.

We now focus on the two-front case. Luckily, the asymptotic effect of the obstacle, $R^{(2)}$, is somewhat simplified in that case. $R^{(2)}$ contains two contributions, one involving $u$ alone, and one involving $u_r^{(1)}$. There is no closed form expression for the contribution that involves $u_r^{(1)}$. However, a detailed inspection reveals that, in the two-front case, this contribution vanishes exponentially away from the origin in all directions. Hence, for the asymptotic effect of the obstacle, only the part of $R^{(2)}$ that is a differential polynomial in $u$ is relevant. Its structure is:

$$R^{(2)} = A R_{31} + B u(t,x) R_{21} + C q(t,x) \left( R_{31} - 2 u(t,x) R_{21} \right)$$

(2.35)

In Eq. (2.35),

$$A = 6 \alpha_1^2 - 9 \alpha_1 \alpha_2 - 3 \alpha_1 \alpha_3 - 3 \alpha_1 \alpha_4 + 12 \alpha_2 \alpha_4 + 3 \alpha_3 \alpha_4 - 6 \alpha_4^2 \\ - 4 \beta_1 - 8 \beta_2 + 4 \beta_3 - 12 \beta_4 + 4 \beta_6 - c \gamma_{21}$$,

$$B = -2 \alpha_1^2 + \tfrac{19}{2} \alpha_1 \alpha_2 - \tfrac{1}{2} \alpha_2^2 + \tfrac{11}{2} \alpha_1 \alpha_3 - \tfrac{7}{2} \alpha_2 \alpha_3 - 5 \alpha_3^2 - 5 \alpha_1 \alpha_4 - 11 \alpha_2 \alpha_4 + 8 \alpha_3 \alpha_4 + 4 \alpha_4^2 \\ + 2 \beta_1 - 10 \beta_2 - 2 \beta_3 + 16 \beta_4 - 6 \beta_6 + 2 c \gamma_{21}$$,

$$C = -\left( \alpha_1 - 2 \alpha_2 - \alpha_3 + 2 \alpha_4 \right) \gamma_{21}$$.

(2.36)

$R_{nm}$, the building blocks for our "canonical" obstacles are given by Eq. (2.9).

Due to Eq. (2.21), $R_{nm}$ vanish in the case of the single-front zero-order solution. In the case of a multi-front zero-order solution, they are sizable only in the interaction region of the fronts. Owing to the unbounded nature of $q(t,x)$, the term proportional to $q(t,x)$ in Eq. (2.35) grows indefinitely and does not lead to a simple asymptotic behavior of $u_r^{(2)}$, unless $C$ of Eq. (2.36) vanishes. This happens only for a specific form of the perturbation in Eq. (2.11), for which either $\gamma_{21} = 0$, or $\left( \alpha_1 - 2 \alpha_2 - \alpha_3 + 2 \alpha_4 \right) = 0$ holds. Of the two possibilities, the first, $\gamma_{21} = 0$, corresponds to the absence of the first-order obstacle. If $C = 0$, a simple asymptotic expression for $u_r^{(2)}$ is obtained. The reason is that $R_{31}$ is proportional to $R_{21}$, (see the Appendix):

$$R_{31} = \left( k_1 + k_2 \right) R_{21}$$

(2.37)

Exploiting Eq. (2.30), the asymptotic solution of Eq. (2.34) is found to be:

$$u_r^{(2)}(t,x) \rightarrow -\tfrac{1}{2} \left( k_1 + k_2 \right) a - \tfrac{1}{2} b u(t,x)$$

(2.38)

 

The analysis in higher orders follows a similar pattern.  We write the $n$'th-order term in the NIT as a sum of two contributions:

$$u^{(n)} = u_s^{(n)}[u] + u_r^{(n)}(t,x)$$

(2.39)

In Eq. (2.39), $u_s^{(n)}[u]$ is the differential polynomial that solves the $n$'th-order homological equation in the case of a single-front solution of the NF.   Consequently, it cancels in the general case all the terms in the homological equation except for those that vanish identically for the single-wave solution.   They constitute the $n$'th-order obstacle, $R^{(n)}$.  $u_r^{(n)}(t,x)$ depends explicitly on $t$ and $x$.  It enables one to account for $R^{(n)}$ through the following equation:

$$\partial_t u_r^{(n)} = 2u_x u_r^{(n)} + 2u \partial_x u_r^{(n)} + \partial_x^2 u_r^{(n)} + R^{(n)}$$

(2.40)

$R^{(n)}$ is built from two types of contributions.    The first involves terms that contain explicit dependence on $t$ and $x$, through $u_r^{(k)}(t,x)$, $k < n$, which have been computed in previous orders.  It is possible that they may be only obtained numerically.  Occasionally, their asymptotic structure may be obtained in closed form.  The second contribution involves the "canonical" obstacles $R_{nm}$ of Eq. (2.9):

$$R^{(n)}[u] = \sum \gamma_{pq}^{(n)} f_{pq}^{(n)}[u] R_{pq}[u]$$

(2.41)

In Eq. (2.41), $\gamma_{pq}^{(n)}$ are known constant coefficients, and $f_{pq}^{(n)}[u]$ are differential polynomials in $u$ that ensure that the obstacle has the correct weight [6, 8, 9].

As in the first-order analysis, all "canonical" obstacles vanish identically in the case of the single-wave solution, as a direct consequence of their form, Eq. (2.9), and of the fact that all symmetries and their integrals are then proportional to $S_1$ and $G_1$, respectively.  In the two-front case, for $k_1 \cdot k_2 < 0$ ($k_1 \cdot k_2 > 0$), all canonical obstacles vanish exponentially for $t \ll 0$ ($t \gg 0$) and become proportional to $u_x$ along the line $x = -(k_1+k_2) \cdot t$, for $t >> 0$ ($t << 0$), because they are all proportional to $R_{21}$. This will be shown in the Appendix.

## 3. NF analysis of the perturbed heat diffusion equation

Consider the perturbed heat-diffusion equation, which we analyze through first-order:

$$w_t = w_{xx} + \varepsilon \left( \alpha_1 w^2 w_x + \alpha_2 w w_{xx} + \alpha_3 w_x^2 + \alpha_4 w_{xxx} \right) + O(\varepsilon^2) \tag{3.1}$$

Such an equation may appear in heat diffusion problems when the conductance depends weakly on temperature. Unlike the case of the Burgers equation, where the 0'th member of the hierarchy was trivial, $S_0 = 0$ and $G_0 = constant$, here there is a nontrivial 0'th member in the hierarchy:

$$S_n = \partial_x{}^n u \qquad\qquad G_n = \partial_x{}^{n-1} u \qquad \left( G_0 = q \right) \tag{3.2}$$

In Eq. (3.2)

$$q \equiv \int\limits_{-\infty}^{x} u(x,t)\,dx \tag{3.3}$$

Through $O(\varepsilon)$, the NF is the heat-diffusion equation with a linear dispersion term:

$$u_t = u_{xx} + \varepsilon \, \alpha_4 \, u_{xxx} + O(\varepsilon^2) \tag{3.4}$$

Expanding $w$ in an NIT, as in Eq. (2.12), the first-order homological equation is

$$u_t^{(1)} = \left[ u_{xx}, u^{(1)} \left[ u; t, x \right] \right] + \alpha_1 u^2 u_x + \alpha_2 u u_{xx} + \alpha_3 u_x^2 \tag{3.5}$$

where the fact that $u^{(1)}$ depends on $t$ and $x$ explicitly is exposed. An obstacle emerges in this order if the explicit dependence on $t$ and $x$ is not included. To see this, we begin with the <u>single-wave solution</u>, for which no obstacle arises:

$$u(t,x) = A \exp\left[k\left(x - v\,t\right)\right] \qquad v = -k - \varepsilon\,\alpha_4\,k^2 + O\!\left(\varepsilon^2\right) \quad .$$ 
(3.6)

As no obstacle emerges in the single-wave case, a differential polynomial solution for $u_s^{(1)}$ is possible. The most general form of $u_s^{(1)}$ that solves the first-order homological equation is:

$$u_s^{(1)} = \left(a + bq\right)u^2 - \left(b + \tfrac{1}{6}\alpha_1\right)q^2\,u_x - \left(a + \tfrac{1}{2}\alpha_2 + \tfrac{1}{2}\alpha_3\right)q\,u_x + c\,u_x \quad .$$ 
(3.7)

For the analysis of the general case, (i.e., when the solution of the NF is not a single-wave one), we adopt the algorithm of Eq. (2.25) for $u^{(1)}$, with $u_s^{(1)}[u]$ of Eq. (3.7), $u$ now being the solution of the NF, Eq. (3.4), in the general case. Substituting the resulting NIT and Eq. (3.4), in Eq. (3.1), the first-order homological equation is found to be

$$\partial_t u_r^{(1)} = \partial_x^2 u_r^{(1)} + R^{(1)} \quad ,$$ 
(3.8)

where the first-order obstacle is given by

$$R^{(1)} = \left(2b + \tfrac{2}{3}\alpha_1\right)u\,R_{02} + \left(2a + \alpha_3 + 2bq\right)R_{12} \quad .$$ 
(3.9)

In Eq. (3.9), $R_{nm}$ are given by Eq. (2.9). As $a$ and $b$ are free, one may simplify the expression for $R^{(1)}$ by choosing $a = -\tfrac{1}{2}\alpha_3$, $b = 0$, to obtain

$$R^{(1)} = \tfrac{2}{3}\alpha_1\,u\,R_{02} \quad .$$ 
(3.10)

Thus, no obstacle exists only if $\alpha_1 = 0$.

In general, Eq. (3.8) may have to be solved numerically for $u^{(1)}$. However, it can be solved in closed form in the case of <u>double-wave solution</u> of the NF, (Eq. 3.4),

$$u(t,x) = A_1 \exp\left[k_1\left(x - v_1 t\right)\right] + A_2 \exp\left[k_2\left(x - v_2 t\right)\right] \qquad v_i = -k_i - \varepsilon\,\alpha_4\,k_i^2 + O(\varepsilon^2) \tag{3.11}$$

if $u^{(1)}$ is allowed to depend on $t$ and $x$ explicitly. (A real harmonic solution is obtained for $A_2 = A_1^*$, $k_1 = i\kappa$, $k_2 = -i\kappa$.) Moreover, even the representation of $u^{(1)}$ as a sum of two terms (Eq. (2.25)) is not required. The reason is that the obstacle of Eq. (3.9) is a linear combination with constant coefficients of the two exponentials

$$\exp\left[\left(2k_1 + k_2\right)x - \left(2k_1 v_1 + k_2 v_2\right)t\right] \quad , \qquad \exp\left[\left(k_1 + 2k_2\right)x - \left(k_1 v_1 + 2k_2 v_2\right)t\right] .$$

If neither of the exponentials is a solution of the homogeneous part of Eq. (3.8), that is, $(k_1 \cdot k_2) \neq 0$, $k_1 \neq -2k_2$ and $k_2 \neq -2k_1$, then Eq. (3.8) is solved in closed form by same linear combination, with the exponentials replaced by

$$-\frac{\exp\left[\left(2k_1 + k_2\right)x - \left(2k_1 v_1 + k_2 v_2\right)t\right]}{\left(2k_1 + k_2\right)^2 + \left(2k_1 v_1 + k_2 v_2\right)} , \qquad -\frac{\exp\left[\left(k_1 + 2k_2\right)x - \left(k_1 v_1 + 2k_2 v_2\right)t\right]}{\left(k_1 + 2k_2\right)^2 + \left(k_1 v_1 + 2k_2 v_2\right)} .$$

(Note that if Eq. (3.11) is replaced by an integral over a continuous range of wave numbers, then a small denominator problem arises in an attempt at a closed-form solution of Eq. (3.8), and it may have to be solved numerically.)

Although $u^{(1)}$ can be solved for in closed form if it is allowed to depend explicitly on $t$ and $x$, we mention here an alternative solution method, which may be of physical interest. Using Eqs. (2.33), (3.2) and (3.3), for a two-wave solution, the canonical obstacles, $R_{nm}$, become

$$R_{nm} = A_1 A_2 \exp\left[\left(k_1 + k_2\right)x - \left(k_1 v_1 + k_2 v_2\right)t\right]\left(k_1 - k_2\right)\left(k_1^n k_2^m - k_1^m k_2^n\right) . \tag{3.12}$$

Consider the choice of Eq. (3.7) for $u_s^{(1)}[u]$ and the simplification that yields the obstacle in the form of Eq. (3.10).  From Eq. (3.12), one concludes that the canonical obstacle, $R_{02}$, vanishes in the anti-symmetric configuration, $(k_2 = -k_1)$, which covers the case of a real oscillatory wave. Consequently, $R^{(1)}$ vanishes then and $u_r^{(1)}[t,x] = 0$.  To analyze the problem in the general situation, when the zero-order approximation is not anti-symmetric $(k_2 \neq -k_1)$, one can perform a Galilean transformation to a moving frame, in which the wave numbers obtain anti-symmetric values, so that the effect of the obstacle can be eliminated.  One solves the problem in the transformed reference frame without obstacles, and then performs the inverse transformation, to obtain the solution of the original problem.

## 4. Concluding remarks

In this paper we have shown how obstacles to integrability may be accounted for in the cases of the perturbed Burgers and heat diffusion equations, by allowing the higher-order corrections in the NIT to depend explicitly on $t$ and $x$.  The NIT ceases to be a sum of differential polynomials in the zero-order approximation (the solution of the NF); parts of it may have to be computed numerically.  The gain is that the NF remains integrable, as it is constructed from symmetries only, and the zero-order term retains the wave structure of the unperturbed solution.  Moreover, the result obtained already in the standard NF analysis, but is not self-evident there, that the obstacles vanish in the single-wave case, is borne out explicitly by the "canonical" form of the obstacles, generated by the algorithm of Eq. (2.39).

In the case of the perturbed Burgers equation, for multi-wave solutions of the NF, the canonical obstacles are sizeable only in the region of interaction among the wave.  Due to the non-localized nature of the solutions, the interaction region extends along a semi-infinite line in the $x$-$t$ plane. As the obstacles fall off exponentially towards zero away from that region, they affect the NIT only within this region.  This leads to a closed-form expression for the asymptotic behavior of the first-order term in the NIT.  This simplicity may not repeat itself in higher orders.

For both the perturbed Burgers and heat diffusion equations, the canonical obstacles obey Eq. (2.56) in the anti-symmetric configuration ($k_2 = -k_1$) of the two-wave zero-order approximation. This result suggests a possible way to eliminate some or all of the obstacles for a two-wave solution of the NF: (1) Transform the given problem to an anti-symmetric one using a Galilean transformation; (2) Exploit whatever freedom is available in the construction of the NIT so that as many obstacles as possible vanish for the anti-symmetric solution; (3) Transform the result to the original reference frame by the inverse transformation.

These systems have the advantage that an obstacle emerges already in the first-order analysis, making the computation a lot easier than in equations that have conservation laws, such as the KdV or NLS equations, for which the zero-order solution is localized (solitons), and an obstacle appears only in second order.  Thus, one may regard the results of the analysis of the systems examined here as guidelines to what may be expected in the application of our algorithm to other perturbed evolution equations.  In addition, they may provide hints for modifying the structure of the generators employed in the analysis in terms of Lie group symmetries (which does not suffer from the disease discussed above) so as to avoid obstacles to integrability.

The shortcoming of the perturbative analysis of the perturbed Burgers equation (and the perturbed heat diffusion equation if exponential solutions with real wave numbers, $k$, are considered) is that, owing to the non-localized nature of the wave solutions of the NF, unbounded terms appear already in the first-order correction (unrelated to whether obstacles to integrability do or do not exist), limiting the validity of the perturbative approximation to $t$ and $x$ of $O(1)$.

**Appendix: Wave number representation**

The properties of canonical obstacles, used in the previous Sections, can be found in a few ways. They are especially easy to obtain through a wave number representation of the solution of the NF. The motivation is the Lax pair representation of the unperturbed Burgers equation [11, 39–42], which involves an auxiliary function $y(t,x)$ related to $u(t,x)$ by the Hopf-Cole transformation

$$y_x = u \, y \quad , \tag{A.1}$$

and the diffusion equation (which is the linearization of the Burgers equation)

$$y_t = y_{xx} \quad . \tag{A.2}$$

Turning to Eq. (2.11), Eq. (A.2) is replaced by the linearization of the NF, Eq. (2.13):

$$y_t = y_{xx} + \varepsilon \, \alpha_4 \, y_{xxx} + \varepsilon^2 \, \beta_6 \, y_{xxxx} + \cdots \quad . \tag{A.3}$$

Writing for $y$

$$y(t,x) = \int A(k) \exp\big(k\big(x - v(k)\,t\big)\big) dk \quad , \tag{A.4}$$

$u$ becomes

$$u(t,x) = \frac{\int k \, A(k) \exp\big(k\big(x - v(k)\,t\big)\big) dk}{\int A(k) \exp\big(k\big(x - v(k)\,t\big)\big) dk} \quad . \tag{A.5}$$

A few words of caution are required. First, to avoid blow up of the solution at a finite point in the ($t$-$x$) plane, $A(k)$ must all have the same sign, which we choose to be positive. Second, if the integration over $k$ is confined to positive (or negative) $k$, then, for sufficiently well behaved $A(k)$, the representation may converge over a whole quadrant in the ($t$-$x$) plane. Otherwise, the integrals may converge only over a finite domain in the plane.  With these words of caution, we note that, with $A(k) > 0$, $u(t,x)$ may be interpreted as the $t$- and $x$- dependent average wave number,

$$u(t,x) = \langle k \rangle \qquad ,$$

(A.6)

under the "probability distribution density"

$$P(k) = \frac{A(k)\exp\big(k\big(x - v(k)t\big)\big)}{\int A(k)\exp\big(k\big(x - v(k)t\big)\big)dk} \qquad .$$

(A.7)

In a similar manner, one easily deduces

$$u_x = \langle k^2 \rangle - \langle k \rangle^2 \qquad .$$

(A.8)

Exploiting Eqs. (2.4)–(2.6), one proves the following relations

$$G_n = \langle k^n \rangle$$
$$S_n = \partial_x G_n = \langle k^{n+1} \rangle - \langle k \rangle \langle k^n \rangle \qquad .$$

(A.9)

Thus, the "potential", $G_n$ is the $n$'th moment of $k$.

In the unperturbed case, substituting Eq. (A.4) in Eq. (A.2), one finds the dispersion relation:

$$v(k) = -k \qquad .$$

(A.10)

In the perturbed case, Eq. (A.3), it is

$$v(k) = -k - \varepsilon \alpha_4 k^2 - \varepsilon^2 \beta_6 k^3 - \cdots \qquad .$$

(A.11)

Eqs. (A.6) − (A.9) hold also in the perturbed case.  This leads to the following expression for the time dependence of $G_n$, which can be derived from the explicit expression of the moments $\langle k^n \rangle$:

$$\partial_t G_n = G_{n+2} - G_n G_2 + \varepsilon \alpha_4 \left( G_{n+3} - G_n G_3 \right) + \cdots$$
$$= S_{n+3} + R_{n,1} + \varepsilon \alpha_4 \left( S_{n+4} + R_{n+1,1} + R_{n,2} \right) + \cdots \quad . \tag{A.12}$$

Thus, as in the case of the unperturbed equation (see Eq. (2.10)), the canonical obstacles appear even before the full expansion procedure is developed.

Due to Eq. (A.9), the obstacles, $R_{nm}$, can be written as

$$R_{nm} = \left\langle k^{n+1} \right\rangle \left\langle k^m \right\rangle - \left\langle k^{m+1} \right\rangle \left\langle k^n \right\rangle \quad . \tag{A.13}$$

The wave-number representation leads to a particularly concise description in the case of isolated front solutions.  For instance, the single-front solution is given by

$$P(k) = P_0 \, \delta(k) + P_1 \delta(k - k_1)$$
$$P_0 = \frac{1}{1 + A_1 \exp\left(k_1 \left(x - v(k_1)t\right)\right)}$$
$$P_1 = \frac{A_1}{1 + A_1 \exp\left(k_1 \left(x - v(k_1)t\right)\right)} \quad . \tag{A.14}$$

With the same notation, the two-front solution is determined by

$$P(k) = P_0 \, \delta(k) + P_1 \delta(k - k_1) + P_2 \, \delta(k - k_2) \quad . \tag{A.15}$$

For a <u>single-front solution</u> of the NF one has

$$G_n = P_1 k_1^n$$
$$S_n = P_1 \left(1 - P_1\right) k_1^{n+1} \quad . \tag{A.16}$$

Consequently, all $S_n$ ($G_n$) are proportional to $S_1$ ($G_1$), and the canonical obstacles $R_{nm}$ (see Eq. (A.13)) vanish.

For a <u>two-front solution</u> of the NF, using Eq. (A.15), the expression of Eq. (A.8) for $u_x$ becomes

$$u_x = P_1 \left(1 - P_1\right)k_1^2 + P_2 \left(1 - P_2\right)k_2^2 - 2P_1 P_2 k_1 k_2 \quad,$$  (A.17)

and for $R_{nm}$ (Eq. (A.13)) –

$$R_{nm} = P_1 P_2 \left(k_1^{\,n} k_2^{\,m} - k_1^{\,m} k_2^{\,n}\right)\left(k_1 - k_2\right) \quad.$$  (A.18)

From Eq. (A.18) one deduces that all canonical obstacles are proportional to the lowest one, $R_{21}$:

$$R_{nm} = k_1^{\,m-1} k_2^{\,m-1} \frac{k_1^{\,n-m} - k_2^{\,n-m}}{k_1 - k_2} R_{21} \quad.$$  (A.19)

As a result, in the anti-symmetric configuration, $k_2 = -k_1$, one has

$$R_{n+1,m} + R_{n,m+1} = 0 \qquad \text{all } n,m$$

$$R_{nm} = 0 \qquad\qquad \left(n + m\right)\text{even} \quad.$$  (A.20)

The    limits    for    $t > 0$    and    $t < 0$    are    interchanged    for    $k_1 \cdot k_2 > 0$.

Figure captions

<u>Fig. 1</u>   Two-front solution of Burgers equation (se Eq. (2.24)); $A_1 = A_2 = 1$, $k_1 = -1$., $k_2 = 0.5$

a͂  $u(x,t)$; b͂  $u_x(x,t)$; c͂  Canonical obstacle $R_{21}$ (Eq.(13))

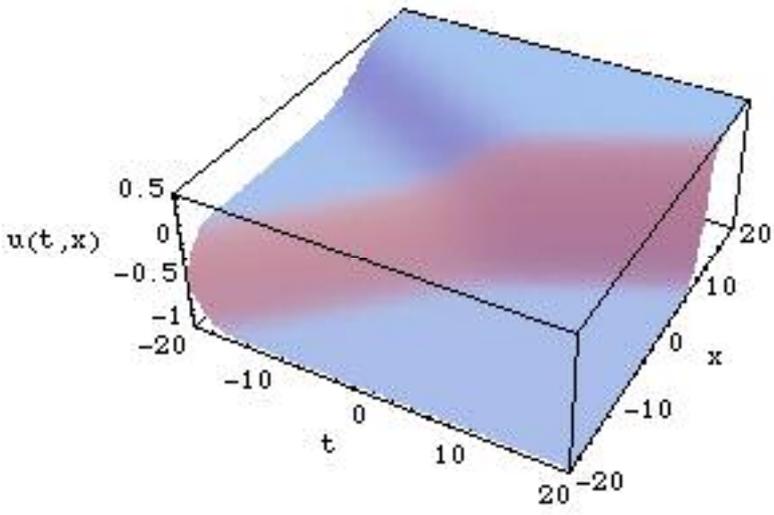

Fig. 1a

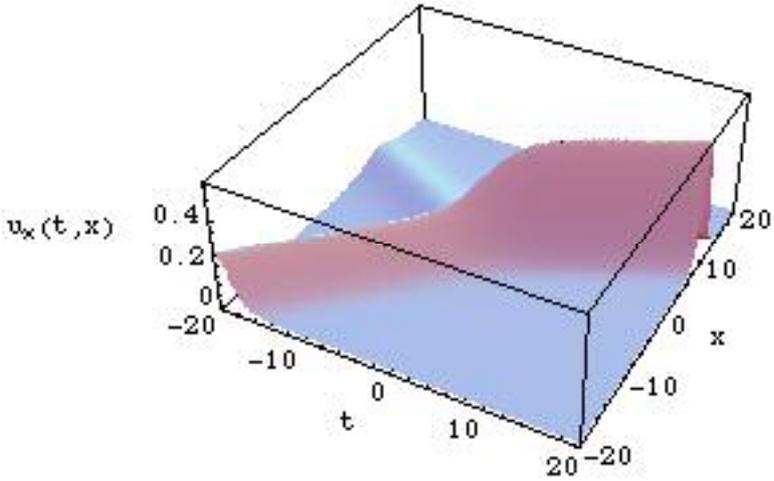

Fig. 1b

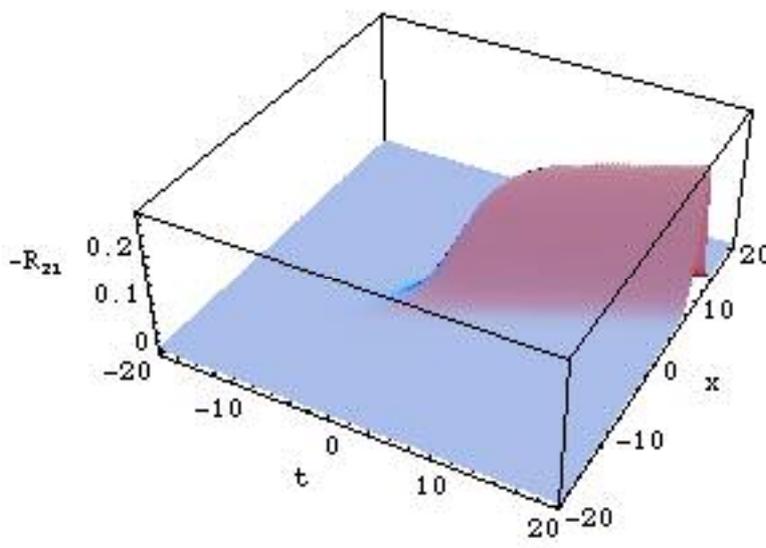

Fig. 1c